\documentclass[10pt,final,journal,letterpaper,oneside,twocolumn]{IEEEtran}

\usepackage[utf8]{inputenc}
\usepackage[T1]{fontenc}
\usepackage{amsmath,amsfonts,amssymb,amsthm,braket}
\usepackage{afterpage}
\usepackage{array,float,stfloats}
\usepackage{booktabs}
\usepackage[justification=justified,singlelinecheck=false,labelfont=bf,labelsep=quad, format=plain,font=small]{caption,subcaption}
\usepackage{changepage}
\usepackage{csquotes}
\usepackage{fancyhdr}
\usepackage{gensymb}
\usepackage{mathtools}
\usepackage{multirow}
\usepackage{newtx}
\usepackage[mode=text]{siunitx}
\DeclareSIUnit\angstrom{\protect \text {Å}}
\usepackage{xcolor, soul}
\sethlcolor{cyan}
\usepackage{graphicx}

\usepackage{subfiles}

\usepackage{xr-hyper}
\usepackage[colorlinks,citecolor=blue,urlcolor=blue,bookmarks=false,hypertexnames=true]{hyperref}
\usepackage[capitalise]{cleveref}

\usepackage{cite}
\usepackage{scalerel}
\usepackage{tikz}
\usetikzlibrary{svg.path}
\definecolor{orcidlogocol}{HTML}{A6CE39}
\tikzset{
  orcidlogo/.pic={
    \fill[orcidlogocol] svg{M256,128c0,70.7-57.3,128-128,128C57.3,256,0,198.7,0,128C0,57.3,57.3,0,128,0C198.7,0,256,57.3,256,128z};
    \fill[white] svg{M86.3,186.2H70.9V79.1h15.4v48.4V186.2z}
                 svg{M108.9,79.1h41.6c39.6,0,57,28.3,57,53.6c0,27.5-21.5,53.6-56.8,53.6h-41.8V79.1z M124.3,172.4h24.5c34.9,0,42.9-26.5,42.9-39.7c0-21.5-13.7-39.7-43.7-39.7h-23.7V172.4z}
                 svg{M88.7,56.8c0,5.5-4.5,10.1-10.1,10.1c-5.6,0-10.1-4.6-10.1-10.1c0-5.6,4.5-10.1,10.1-10.1C84.2,46.7,88.7,51.3,88.7,56.8z};}}
\newcommand\orcidicon[1]{\href{https://orcid.org/#1}{\mbox{\scalerel*{
\begin{tikzpicture}[yscale=-1,transform shape]
\pic{orcidlogo};
\end{tikzpicture}
}{|}}}}

\begin{document}
\title{Variable-temperature attenuator calibration method for on-wafer microwave noise characterization of low-noise amplifiers}

\author{Anthony J. Ardizzi\orcidicon{0000-0001-8667-1208}, Jiayin Zhang, Akim A. Babenko \orcidicon{0000-0002-5552-4771}, Kieran A.\ Cleary\orcidicon{0000-0002-8214-8265}, and Austin J. Minnich\orcidicon{0000-0002-9671-9540}
\thanks{This work was sponsored by the Keck Institute for Space Studies and the National Science Foundation under Grant No.\ 2511983.  (\textit{Corresponding author: Anthony J. Ardizzi})}
\thanks{Kieran A. Cleary is with the Cahill Radio Astronomy Lab, Division of Physics, Mathematics and Astronomy, California Institute of Technology, Pasadena, CA 91125, USA (e-mail: kcleary@astro.caltech.edu)}
\thanks{Akim A. Babenko is with the Jet Propulsion Laboratory, Pasadena, CA 91011, USA}
\thanks{Anthony J. Ardizzi, Jiayin Zhang, and Austin J. Minnich are with the Division of Engineering and Applied Science, California Institute of Technology, Pasadena, CA 91125, USA (e-mail: aardizzi@caltech.edu; jiayinz@caltech.edu; aminnich@caltech.edu)}}

\maketitle

\begin{abstract}
Low-noise cryogenic microwave amplifiers are widely used in  applications such as radio astronomy and quantum computing. On-wafer noise characterization of cryogenic low-noise transistors is desirable because it facilitates more rapid characterization of devices prior to packaging, but obtaining accurate noise measurements is difficult due to the uncertainty arising from the input loss and temperature gradients prior to the device-under-test (DUT). Here, we report a calibration method that enables the simultaneous determination of the backend noise temperature and effective-noise-ratio at the input plane of the DUT. The method is based on measuring the S-parameters and noise power of a series of attenuators at two or more distinct physical temperatures. We validate our method by measuring the noise temperature of InP HEMTs in 4-8 GHz. The calibration method can be generalized to measure the microwave noise temperature of any two-port device so long as a series of attenuators can be measured at two or more distinct physical temperatures.
\end{abstract}



\section{Introduction}
Low-noise microwave amplifiers (LNAs) are of high importance for many applications such as quantum computing \cite{bardin_microwaves_2021,hornibrook_cryogenic_2015}, radio astronomy \cite{pospieszalski_extremely_2018,ajayan_critical_2021,chiong_low-noise_2022}, atmospheric radiometry \cite{reising_inp_2012,cooke_670_2021}, mm-wave imaging security systems\cite{kolinko_passive_2005,sato_inp-hemt_2008,wikner_millimeter_2017}, and ``Beyond 5G'' technology\cite{hamada_millimeter-wave_2019,hamada_ultra-high-speed_2021,tsutsumi_systematic_2024}. Among semiconductor technologies, InP HEMTs exhibit the best noise performance. Increasingly, superconducting parametric amplifiers are finding application for the most demanding applications such as quantum computing. \cite{aumentado_superconducting_2020}


Owing to the outstanding noise performance of cryogenic low-noise amplifiers, which approaches the quantum limit for superconducting amplifiers, characterizing their noise temperature has become increasingly difficult. For such low-noise amplifiers, the measured noise temperature is more sensitive to systematic uncertainties compared to noisier devices. Further, system noise calibration is challenging to perform in a cryogenic vacuum environment due to the complexities of accurately measuring input loss and noise power at the input plane of the DUT. Several approaches have been explored in the literature, most of which employ some version of the Y-factor method\cite{cano_ultra-wideband_2010,gu_measurement_2013,randa_detailed_2016,leffel_y_2021}. Several studies have implemented a variable-temperature-load to sweep source power \cite{bruch_noise_2012,kooi_programmable_2018}, and a shot-noise tunnel-junction noise source was also used to generate noise power \cite{su-wei_chang_noise_2016,zou_cryogenic_2025}. Other implementations use cryogenic impedance tuners to directly measure noise parameters\cite{russell_cryogenic_2012,sheldon_noise-parameter_2021}. In all cases, the noise contribution of both the receiver and input components must be carefully calibrated to extract the noise contribution of the device-under-test (DUT), a problem for which a robust and reliable calibration method remains an open challenge.


\begin{figure}[t]
    \centering
    \includegraphics[width=0.49\textwidth]{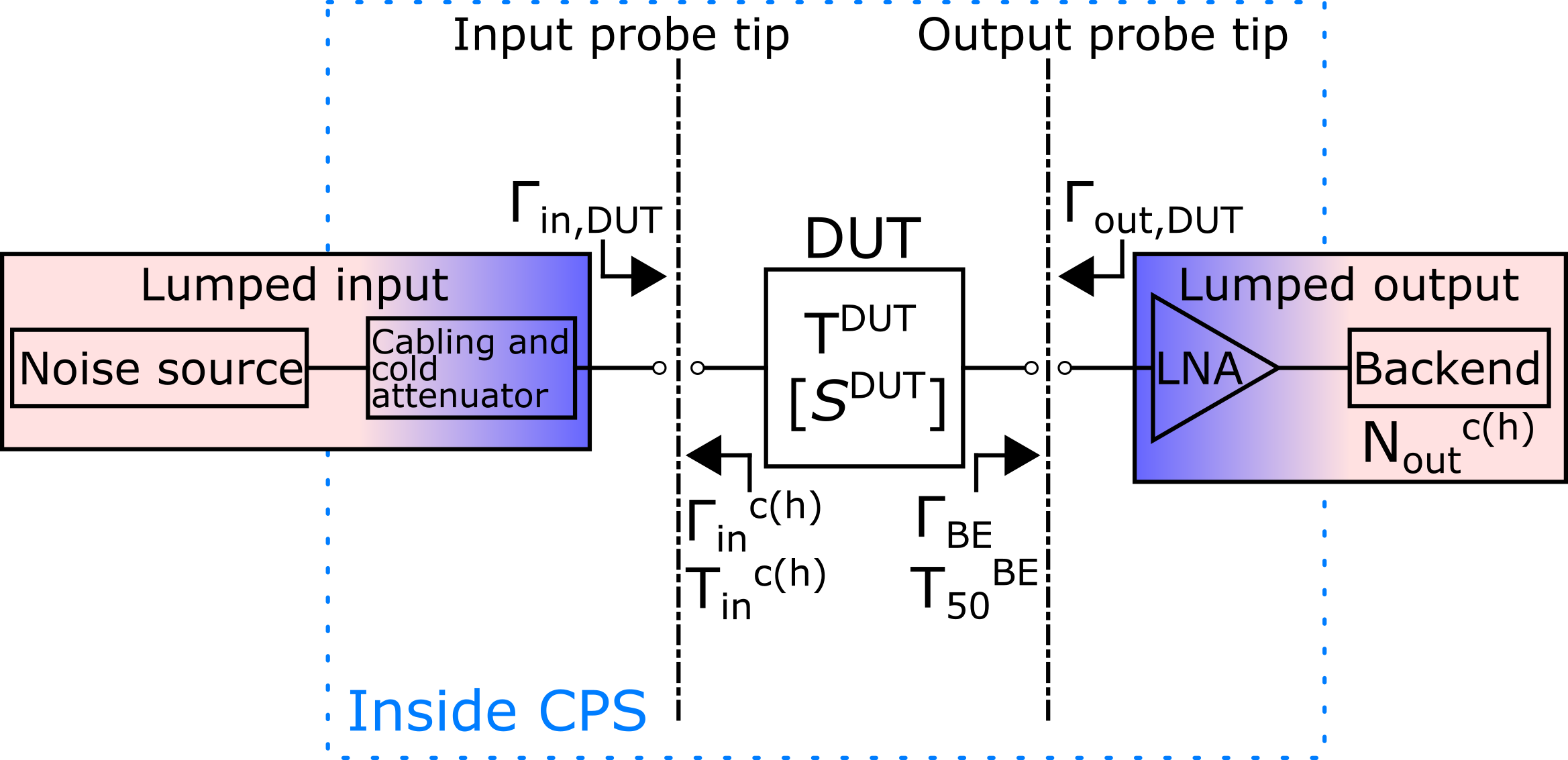}
    \caption{Schematic of the CPS showing all lumped component parameters. The color gradient represents the temperature gradient from room temperature to cryogenic temperatures. The significance of each parameter is described in the text.\label{fig:LumpedSchem}}
\end{figure}

Here, we report a cryogenic noise calibration technique which relies on measuring the noise and S-Parameters of a series of attenuators at two distinct physical temperatures. The backend noise temperature and effective-noise ratio at the input plane of the DUT can be simultaneously extracted from these measurements, providing the data needed to measure the noise of the DUT. We demonstrate the method by measuring the noise temperature of InP HEMTs in 4-8 GHz and comparing the values to those obtained with the traditional calibration method. Our work addresses a long-standing challenge in cryogenic noise measurement and will facilitate accurate noise characterization of low-noise amplifiers.


The article is structured as follows. In \cref{sec:AttenCal}, we introduce the theory underlying the calibration method and demonstrate the concept using electromagnetic simulations. In \cref{sec:CPS}, we describe the experimental approach based on a cryogenic probe station (CPS). We then experimentally validate the calibration procedure and demonstrate its application to InP HEMTs in \cref{sec:HEMTs}. Finally, we conclude in \cref{sec:conclusion}.



\section{On-wafer cryogenic noise calibration \label{sec:AttenCal}}


\begin{figure*}[t]
    {\phantomsubcaption\label{fig:CryoAttenSim}}
    {\phantomsubcaption\label{fig:CryoAttenExp}}
    \centering
    \includegraphics[width=0.85\textwidth]{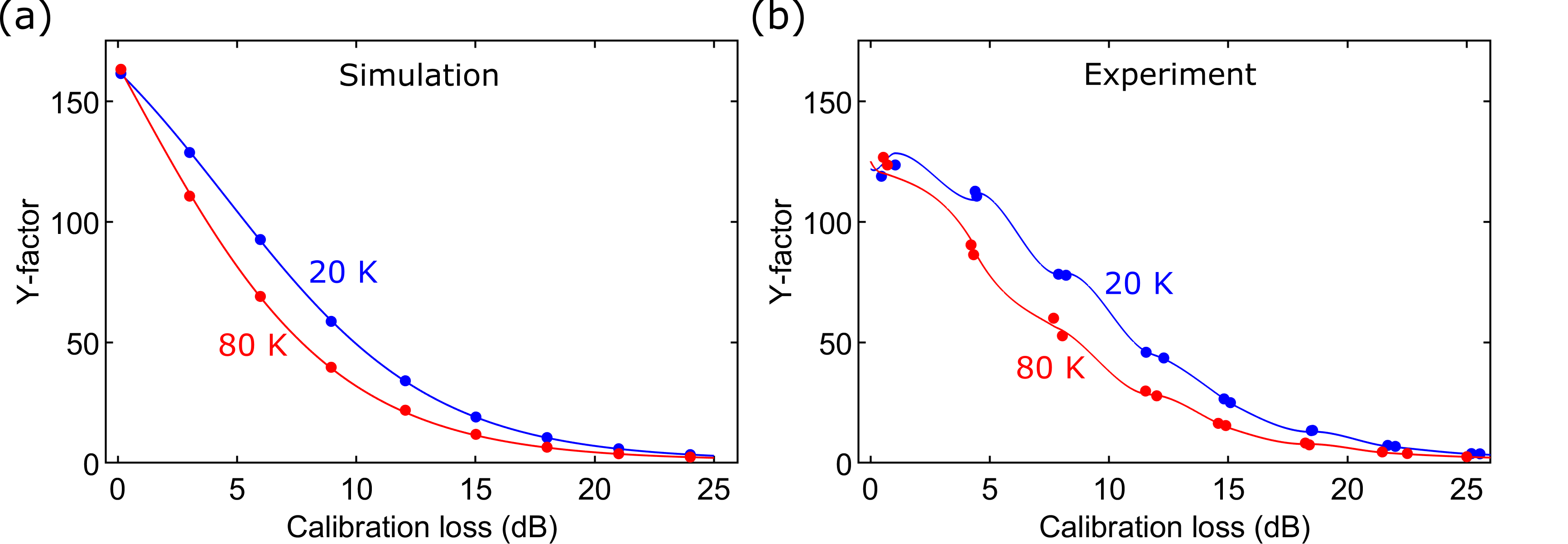}
    \caption{\textbf{(a)} Y-factor vs attenuator loss data simulated assuming perfect impedance matching. Simulated data points at physical temperatures of 20~K (blue) and 80~K (red) are shown. Solid lines show the fit to \cref{eq:YfacSimple} \textbf{(b)} Experimentally measured Y-factor vs attenuator loss. Solid lines show the fit to \cref{eq:YFactAtten} with interpolation between points to account for discrete changes in the correction coefficients. The non-monotonic line-shape is due to changes in impedance mismatch between devices of different losses. Error bars are the size of the symbols.}
\end{figure*}

We begin by describing the theory of the calibration technique based on measurement of  S-parameter and Y-factor measurements of a series of attenuators at two distinct physical temperatures. A schematic of a two-port noise measurement showing the relevant parameters that require calibration is shown in \cref{fig:LumpedSchem}. In particular, the noise source and input cabling are lumped into a single input noise source with noise temperature $T_\text{in}^\text{c(h)}$ and reflection coefficient $\Gamma_\text{in}^\text{c(h)}$ with the noise source off (on). The backend receiver, which includes the output probe tip and LNA, also contributes input-referred noise $T^\text{BE}(\Gamma_\text{out,DUT}^\text{c(h)})$ with reflection coefficient $\Gamma_\text{BE}$. Finally, the DUT has an input-referred noise temperature $T^\text{DUT}(\Gamma_\text{in}^\text{c(h)})$, S-parameter matrix $[S^{DUT}]$, and output reflection coefficient $\Gamma_\text{out,DUT}^\text{c(h)}$. The noise power reaching the detector is $N_\text{out}^\text{c(h)}$ with the noise source off (on).

Following the standard noise propagation procedure (c.f. Ref. \cite{pozar_microwave_2012} Ch.10), the Y-factor equation $Y\equiv N_\text{out}^\text{h}/N_\text{out}^\text{c}$ giving the ratio of hot power to cold power measured at the detector can be obtained as:

\begin{equation}
    \label{eq:YFactGen}
    Y=\frac{(T_\text{in}^\text{h} + T^\text{DUT}(\Gamma_\text{in}^\text{h}))G_\text{a,DUT}^\text{h} G_\text{a,BE}^\text{h} + T^\text{BE}(\Gamma_\text{out,DUT}^\text{h}) G_\text{a,BE}^\text{h}}{(T_\text{in}^\text{c} + T^\text{DUT}(\Gamma_\text{in}^\text{c}))G_\text{a,DUT}^\text{c} G_\text{a,BE}^\text{c} + T^\text{BE}(\Gamma_\text{out,DUT}^\text{c}) G_\text{a,BE}^\text{c}}
\end{equation}
where $G_\text{a,i}^\text{c(h)}$ represents the available gain of the i$^\text{th}$ component with the noise source off (on). It is important to specify the noise source state since the source impedance typically changes between noise source states for most avalanche diode noise sources commonly used for Y-factor measurements \cite{gu_measurement_2013}, and both the available gain and input-referred noise temperature of active components depend on the impedance of the preceding components in the network. In terms of reflection coefficients and S-parameters, the available gain is given by \cite{orfanidis_chapter_2016}:

\begin{equation}
    \label{eq:AvailableGain}
    G_\text{a,i}=\frac{1-|\Gamma_\text{out}^{i-1}|^2}{|1-S_{11}^i\Gamma_\text{out}^{i-1}|^2}|S_{21}^i|^2\frac{1}{1-|\Gamma_\text{out}^i|^2}
\end{equation}
and the input-referred noise temperature $T_\text{i}$ of an active device ($G_\text{a}>1$) is given by the noise parameter equation \cite{randa_detailed_2016}:
\begin{equation}
    \label{eq:Tnoiseparams}
    T_\text{i}(\Gamma_\text{out,i-1})=T_\text{min} + \frac{4R_NT_0}{Z_0}\frac{|\Gamma_\text{out,i-1}-\Gamma_\text{opt}|^2}{(1-|\Gamma_\text{out,i-1}|^2)|1+\Gamma_\text{opt}|^2}
\end{equation}
where $\Gamma_\text{out,i-1}$ is the output reflection coefficient of the source network, $T_\text{min}$ is the minimum noise temperature of the DUT, $R_\text{N}$ is the noise resistance, $T_0=290$~K is the reference noise temperature, $Z_0=50~\Omega$ is the reference impedance, and $\Gamma_\text{opt}$ is the optimum reflection coefficient for noise matching.

The above set of equations represents the most general Y-factor formulation which accounts for all impedance mismatches in the measurement chain. To simplify this equation for the calibration procedure, in which case the DUT is a passive component ($G_\text{a,DUT}<1$) which we designate as "cal", we exploit that the physical temperature $T_\text{phys}$ of the component is related to its input-referred noise temperature $T^\text{DUT}$ through the equation $T_\text{phys}=T^\text{DUT}G_\text{a,DUT}/(1-G_\text{a,DUT})$ \cite{pettai_noise_1984} to rewrite the Y-factor equation for an attenuator as:

\begin{equation}
    \label{eq:YFactAtten}
    Y=\mathbb{C}\frac{(T_\text{in}^\text{h}-T_\text{phys}^\text{cal})G_\text{a,cal}^\text{h}+(T_{e}^\text{BE}+T_\text{phys}^\text{cal})}{(T_\text{in}^\text{c}-T_\text{phys}^\text{cal})G_\text{a,cal}^\text{c}+(T_{e}^\text{BE}+T_\text{phys}^\text{cal})}
\end{equation}

Here, $\mathbb{C}$ is a correction coefficient accounting for the impedance match between the attenuator and the backend receiver:

\begin{align}
    \label{eq:CorCoeffs}
    \mathbb{C}&= \frac{|1-S_{11}^\text{BE}\Gamma_\text{out,cal}^\text{c}|^2(1-|\Gamma_\text{out,cal}^\text{h}|^2)}{|1-S_{11}^\text{BE}\Gamma_\text{out,cal}^\text{h}|^2(1-|\Gamma_\text{out,cal}^\text{c}|^2)}
\end{align}

The mismatch between the attenuator and the noise source is taken into account by using the available gain. Both $|S_{21}^\text{BE}|^2$ and $1/(1-|\Gamma_\text{out}^\text{BE}|^2)$ cancel between numerator and denominator; the former because it does not depend on impedance change, and the latter because the product $S_{21}^\text{BE}S_{12}^\text{BE}\ll1$ so that $\Gamma_\text{out}^\text{BE}=S_{22}^\text{BE}$ which again does not depend on impedance change. The assumption that $S_{21}^\text{BE}S_{12}^\text{BE}\ll1$ is satisfied in our setup since the active directivity of all backend amplifiers in our setup are much greater than 1. This requirement can be further guaranteed by the use of an isolator, but we found this to be unnecessary. The reflection coefficients are defined as shown in \cref{fig:LumpedSchem}, with the additional specification that $\Gamma_\text{out,cal}^\text{c(h)}$ is defined with the attenuator input terminated in the lumped input impedance with the noise source off (on).

We note that we have assumed that the backend noise temperature does not change with input impedance; that is $T^\text{BE}(\Gamma_\text{out,DUT}^\text{c})\approx T^\text{BE}(\Gamma_\text{out,DUT}^\text{h})\equiv T_{e}^\text{BE}$. While this assumption is not strictly satisfied in general, it is often a good approximation because the change in input-network impedance as seen by the backend will be attenuated by the DUT loss. A more precise method would be to independently measure the noise parameters of the backend so that $T^\text{BE}(\Gamma)$ is known for all $\Gamma$, but this approach requires substantially more measurements. 

\Cref{eq:YFactAtten} provides a method to determine the three quantities that must be calibrated for noise measurements of active devices, namely $T_\text{in}^\text{c}$, $T_\text{in}^\text{h}$, and $T_{50,T_\text{phys}}^\text{BE}$. By measuring the Y-factor and S-parameters of a series of attenuators, as well as the reflection coefficients of the lumped input and output, these quantities can in principle be extracted from a fit to \cref{eq:YFactAtten}. Before demonstrating this, we first verify whether the fitting function is well-conditioned by considering the simplest case where all components are perfectly matched, so that $\mathbb{C} = 1$. Under these conditions, we can simplify the Y-factor equation to:

\begin{align}
\label{eq:YfacSimple}
    Y =& \frac{(T_\text{in}^\text{h} - T_\text{phys})|S_{21}^\text{cal}|^2 + (T^\text{BE}_{50} + T_\text{phys})}{(T_\text{in}^\text{c} - T_\text{phys})|S_{21}^\text{cal}|^2 + (T^\text{BE}_{50} + T_\text{phys})}\\
    =& \frac{Ax + 1}{Bx + 1}
\end{align}
where $x\equiv|S_{21}^\text{cal}|^2$, $A\equiv(T_\text{in}^\text{h} - T_\text{phys})/(T^\text{BE}_{50} + T_\text{phys})$ and $B \equiv (T_\text{in}^\text{c} - T_\text{phys})(T^\text{BE}_{50} + T_\text{phys})$.

When written in this form, two important facts become apparent. The first is that if either $T_\text{in}^\text{h}=T_\text{phys}$ or $T_\text{in}^\text{c}=T_\text{phys}$, the Y-factor becomes insensitive to that quantity, meaning that using this technique at room temperature is not viable without either heating of the attenuator or the use of a noise source emitting both hot and cold noise at a temperature sufficiently different from room temperature. The second is that there are only two independent fitting parameters, indicating that only two of the three required calibrations can be performed.

\begin{figure}[t]
    {\phantomsubcaption\label{fig:AttenPic}}
    {\phantomsubcaption\label{fig:AttenSPars}}
    \centering
    \includegraphics[width=0.48\textwidth]{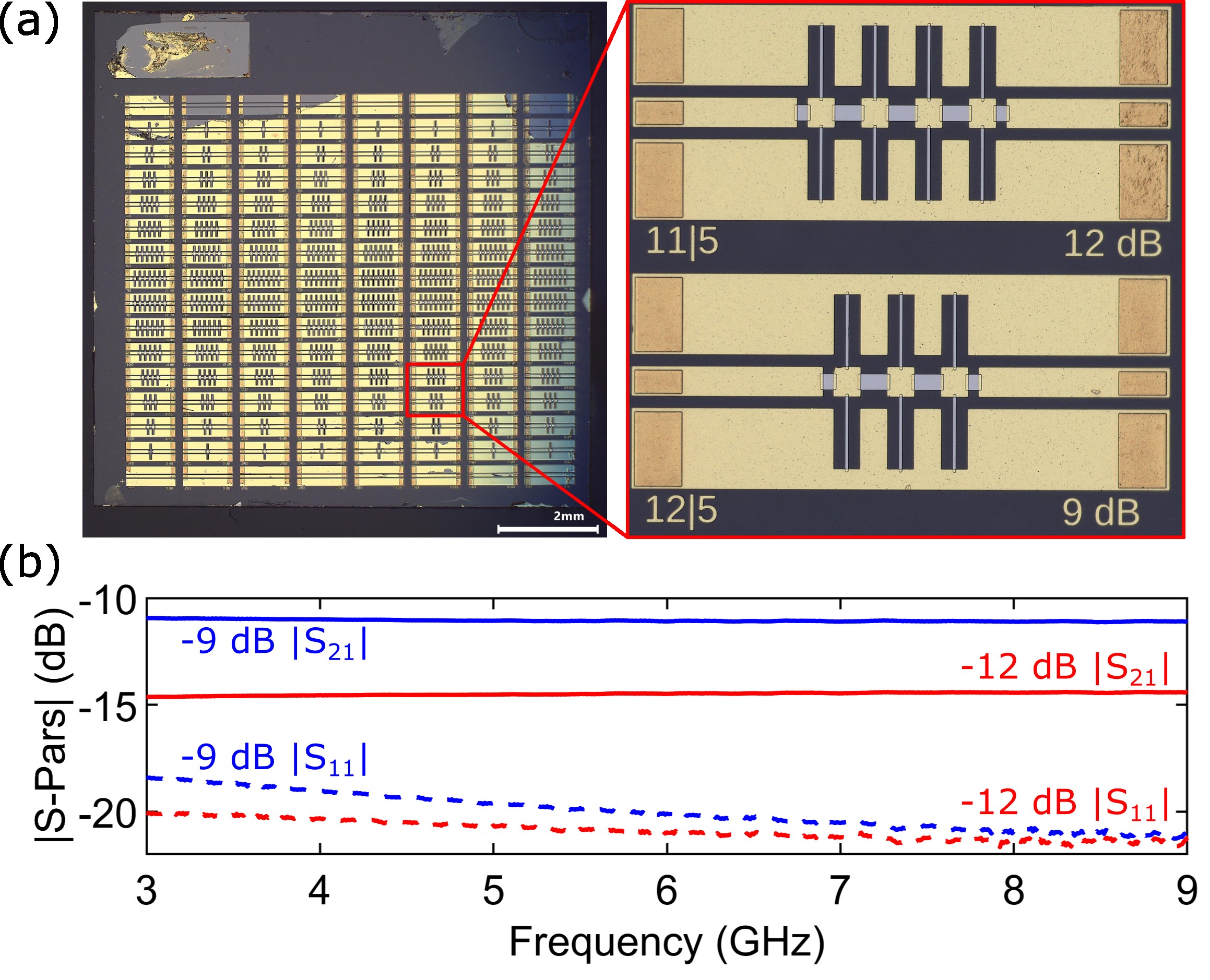}
    \caption{\textbf{(a)} Microscope image of the attenuator chip used in this study. Zoomed image shows two characteristic devices. \textbf{(b)} S-parameters of pictured devices. The measured loss differs from the design by a few dB, which is attributed to a thinner than intended NiCr layer.}
\end{figure}

To overcome this limitation, a second set of measurements at another distinct physical temperature can be used to fit a set of four independent parameters, namely $T_\text{in}^\text{c(h)}$, $T_{50,T_\text{phys,1}}^\text{BE}$, and the backend noise temperature at the two distinct physical temperatures $T_{50,T_\text{phys,1}}^\text{BE}$ and $T_{50,T_\text{phys,2}}^\text{BE}$, where it is implicitly assumed that the noise from the lumped input does not change between physical temperatures. In practice, this is challenging to achieve without decoupling the attenuator chip heating from the rest of the system. For our proof-of-principle experiment where the attenuator temperature was set by heating the entire cold-stage, we ensured that all input components including coaxial cabling, attenuator, bias tee, and probes were well thermalized directly to the cold fingers of the cryogenic system, such that their temperature was nearly independent of the stage temperature. While the probe tips will heat due to their direct contact with the devices mounted to the stage, their loss is sufficiently small that such heating can in general be neglected at all cryogenic temperatures considered here. We emphasize again that the optimal implementation of this technique is to use a floating chip similar to that presented in \cite{kooi_programmable_2018}, but with an array of switchable attenuators.

To demonstrate the viability of this technique, Keysight Advanced Design System (ADS) was first used to simulate the Y-factor data from our probe station at physical temperatures of 20~K and 80~K and 5~GHz. For simplicity in this proof-of-concept demonstration, perfect impedance match between components was assumed, and the hot and cold noise reaching the attenuator was taken to be 50000~K and 200~K, respectively, to approximately reflect our experimental setup for ungated HEMTs (with no cryogenic attenuator present). The backend noise temperature was set to $T_{50,T_\text{phys}}^\text{BE}=60$~K at both physical temperatures. The final noise powers $N_\text{out}^\text{c(h)}$ were exported into Matlab, and artificial Gaussian noise was added to both the noise powers ($\sigma/\mu=0.5~\%$) and attenuator loss ($\sigma/\mu=0.05$~dB, converted to linear error) to reflect actual experimental uncertainties in each simulated data point. A least squares fit to the simulated noisy Y-factor versus loss data was then performed using the Matlab algorithm \textit{lsqcurvefit}. 

\begin{figure*}[t]
    {\phantomsubcaption\label{fig:CPS}}
    {\phantomsubcaption\label{fig:Backend}}
    {\phantomsubcaption\label{fig:CPSpic}}
    \centering
    \includegraphics[width=0.90\textwidth]{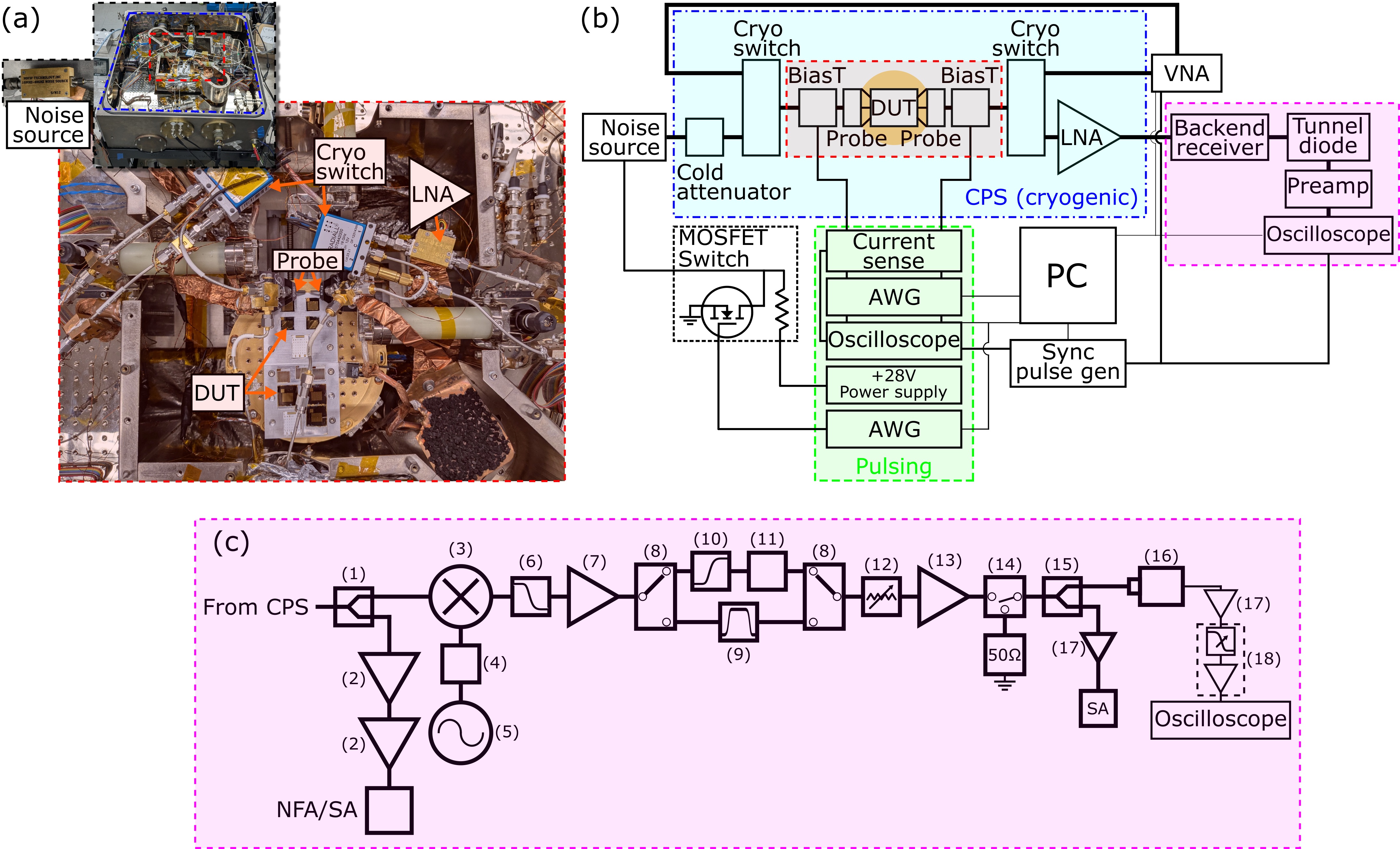}
    \caption{\textbf{(a)} Schematic depiction of the CPS, including relevant electronics. \textbf{(b)} Schematic of the room temperature backend receiver used for noise measurements. The numbered components are listed in \cref{tab:RTbackend}. \textbf{(c)} Picture of the CPS with relevant electronics and microwave components in focus. The backend receiver is thermally insulated within a box to minimize 1/f noise from ambient environment fluctuations.\label{fig:CPSfull}}
\end{figure*}

The simulation results are shown in \cref{fig:CryoAttenSim}, which shows good agreement between simulated data and the fit, as expected. The extracted calibration parameters are shown in \cref{tab:SimParams} and were calculated as follows. First, the 95\% confidence interval in the four fit parameters $A_{T_1}$, $A_{T_2}$, $B_{T_1}$, $B_{T_2}$ was computed by the Matlab algorithm \textit{nlparci}. The 95\% confidence interval was then divided by the number of standard errors spanning this interval as computed by the function \textit{tinv} for 14 degrees of freedom (18 measurements less 4 fit parameters), which evaluates to $\sim2.14$. Finally, these errors were propagated to the calibration parameters of interest through common propagation of error techniques \cite{taylor_introduction_1997}.

The extracted values agree within uncertainty to the true values used in the ADS simulation. While it is challenging to directly compare these uncertainties with other measurement setups reported in the literature since $T_\text{in}^\text{c(h)}$ at the input probe tip is not a commonly reported quantity, the percent errors reported here are at least competitive with those previously reported by our group (see the SI of \cite{ardizzi_self-heating_2022} for example). 

\begin{table}[!hb]
\renewcommand{\arraystretch}{1.3}
\begin{minipage}[t]{0.48\textwidth}
\captionsetup{justification=justified}
\caption{Extracted calibration parameters from ADS simulation. \label{tab:SimParams}}
\vspace{-12pt}
\begin{tabular}{| m{0.17\textwidth} | m{0.3\textwidth} | m{0.2\textwidth} |}
\hline
\textbf{Parameter} & \textbf{Extracted value} & \textbf{Simulation value}\\
\hline
$T_\text{in}^\text{c}$ & $271\pm15$~K & 250~K \\
\hline
$T_\text{in}^\text{h}$ & $52172\pm3055$~K & 50000~K \\
\hline
$T_{50,T_\text{phys,1}}^\text{BE}$ & $63.1\pm4.8$~K & 60~K\\
\hline
$T_{50,T_\text{phys,2}}^\text{BE}$ & $63.4\pm8.6$~K & 60~K \\
\hline
\end{tabular}
\vspace{20pt}
\end{minipage}%
\hfill 
\hfill

\begin{minipage}[t]{0.48\textwidth}
\captionsetup{justification=justified}
\caption{Extracted calibration parameters from experiment with no cold-attenuator on the input line. The estimated values were calculated from independently measuring and cascading the noise from each component. \label{tab:ExpParams}}
\begin{tabular}{| m{0.17\textwidth} | m{0.3\textwidth} | m{0.3\textwidth} |}
\hline
\textbf{Parameter} & \textbf{Extracted value} & \textbf{Estimated value}\\
\hline
$T_\text{in}^\text{c}$ & $392\pm18$~K & $324\pm40$~K \\
\hline
$T_\text{in}^\text{h}$ & $60507\pm2478$~K & $64246\pm3632$~K \\
\hline
$T_{50}^\text{BE}$ & $54\pm3$~K & $31\pm9$~K\\
\hline
\end{tabular}
\end{minipage}
\end{table}

To test the calibration method experimentally, on-chip 100~$\mu$m pitch tee attenuators ranging from 0~dB to 21~dB were designed and fabricated on a sapphire substrate. \Cref{fig:AttenPic} shows the chip with an array of attenuators which were used in this study. The NiCr 80/20 resistors were first deposited using electron beam evaporation, followed by the Au metallization layer. Finally, the pads were gold plated to facilitate probing. \Cref{fig:AttenSPars} shows the $|S_{21}|$ and $|S_{11}|$ of the two characteristic attenuators pictured in \cref{fig:AttenPic}. While the return loss is better than -18~dB across the frequency band measured here, the insertion loss is greater than designed by $\sim2$~dB, which we attribute to a thinner NiCr layer than intended. We expect that further fabrication iterations would improve both insertion and return loss relative to design parameters.

This chip was mounted in our cryogenic probe station (CPS) which is shown in \cref{fig:CPSfull} and described in greater detail in \cref{sec:CPS} below. A series of noise and S-parameter measurements were taken at 20~K and 80~K using the ungated-HEMT setup described there. The physical temperature of the stage as measured by a Lakeshore DT-670 temperature diode was used as a proxy for attenuator temperature. The data and best fit using the fully corrected \cref{eq:YFactAtten} are shown in \cref{fig:CryoAttenExp}. The fit is in good quantitative agreement with the measured data as shown by a low normalized RMSE = 0.044, where the normalization was done by dividing the RMSE by mean(Y)~$=45.3$.

The non-monotonic line-shape is caused by the discretely changing correction coefficients between each data point, which were smoothly interpolated between before plotting. The extracted calibration parameters are shown in \cref{tab:ExpParams}. We also further constrained $T_{50,T_\text{phys,1}}^\text{BE}=T_{50,T_\text{phys,2}}^\text{BE}$ to minimize the number of fitting parameters. This assumption is justified given that the backend LNA changes physical temperature by less than 10~K across the temperature range used here, and independent probe station measurements have shown that this change in physical temperature leads to a negligible change in backend noise temperature.

The standard errors reported in \cref{tab:ExpParams} were calculated by performing a Monte Carlo simulation. Sets of input data were generated by sampling from a standard distribution of data whose mean and variance deviation were determined by the measured data and assumed primary uncertainty, respectively. These data sets were then run through the fitting algorithm. We performed 1000 simulations, and the fitted parameter variance was computed from the final distribution. We assumed uncertainties in the physical temperature $\Delta T_{phys}=1$~K, the S-parameters $\Delta [S]=0.1$~dB, and the measured power $\Delta P/P=0.05$. We also assumed that our entire measurement chain, including our square-law power detector, was perfectly linear with input power. We note that the uncertainties computed using this method are of the same order as the extracted uncertainties determined directly from the Matlab fitting algorithm, indicating both that our assumptions of the primary uncertainties were justified, and that our fitting is robust with respect to modest primary calibration uncertainties. The extracted fitting uncertainties are better than 5\%, which is on par with state-of-art calibration techniques (c.f. \cite{fernandez_noise-temperature_1998} Table 1), especially when considering that our technique accounts for all impedance mismatches and yields calibration parameter uncertainties which are directly transferred from a single physical temperature measurement, that of the calibration attenuators, the S-parameter calibration, and the uncertainty in measured power.

\begin{table*}[t]
\captionsetup{justification=centering}
\caption{List of components used in the backend receiver. \label{tab:RTbackend}}
\centering
\begin{tabular}{| m{0.1\textwidth} | m{0.5\textwidth} | m{0.3\textwidth} |}
\hline
\textbf{Component} & \textbf{Model} & \textbf{Description} \\
\hline
(1) & Krytar MLDD 2-way power divider 1-18~GHz & Power splitter \\
\hline
(2) & Mini-Circuits ZX60-14012L-S+ 300~kHz-14~GHz amplifier & Low noise wideband amplifier \\
\hline
(3) & Marki M1-0220-P mixer & Double-balanced mixer \\
\hline
(4) & -5~dB attenuator pad & Fixed attenuator \\
\hline
(5) & Hittite HMC-T2100 Signal Generator & Signal generator \\
\hline
(6) & Mini-Circuits SLP-250+ 0-225~MHz LPF & Low-pass filter \\
\hline
(7) & Mini-Circuits ZFL-500LN+ 500~MHz amplifier & Low noise RF amplifier \\
\hline
(8) & Transco SPDT latching 28~V RF switch 0-18~GHz & RF switch \\
\hline
(9) & Mini-Circuits SBP-21.4+ 19.2-23.6~MHz BPF & Bandpass filter \\
\hline
(10) & Mini-Circuits SHP-25+ 27.5-800MHz HPF & High-pass filter \\
\hline
(11) & Mini-Circuits VAT-10+ -10dB attenuator & Fixed attenuator \\
\hline
(12) & Mini-Circuits ZX73-2500-S 0-40B voltage-variable attenuator & Voltage-variable attenuator \\
\hline
(13) & Mini-Circuits ZFL‐1000H+ 1GHz amplifier & Medium-power amplifier \\
\hline
(14) & Mini-Circuits MSP2TA-18-12+ Absorptive DC-18GHz switch & Dicke-style switch with 50Ohm termination \\
\hline
(15) & Mini-Circuits ZX10-2-12-S+ 2-1200MHz splitter & Power splitter \\
\hline
(16) & Fairview Microwave FMMT6001 Tunnel Diode Zero Bias Detector & Tunnel diode detector \\
\hline
(17) & SRS SR445A four-channel DC-350MHz amplifier & Preamplifier \\
\hline
(18) & SRS SR560 low noise voltage preamplifier & Preamplifier \\
\hline
\end{tabular}
\end{table*}

The fitted values also agree reasonably with the estimated values shown in the third column of \cref{tab:ExpParams}, which were estimated based on an independent noise source calibration, cable loss measurements, estimates of the temperature gradient along the coaxial cabling\cite{soliman_thermal_2016} and cryogenic switches, and the LNA datasheet. The uncertainty in the estimated $T_\text{in}^\text{c}$ primarily reflects the uncertainty in the estimated distribution of temperature and loss along the input and output components, whereas the uncertainty in the estimated $T_\text{in}^\text{h}$ strongly reflects the uncertainty in the noise source calibration (calibrated using the standard liquid nitrogen-cooled fixed load method described elsewhere, see for example section S.3. in the SI of \cite{ardizzi_self-heating_2022}). The uncertainty in $T_{50}^\text{BE}$ reflects the uncertainty in loss and temperature gradient along the output cabling from the output probe tip to the backend LNA.  

We also note that $T_\text{in}^\text{c}>300$~K, which at first seems unphysical since the off-state noise source emits noise at room temperature. This discrepancy is resolved by considering that $T_\text{in}^\text{c}$ is defined as the temperature that a fictitious resistor, with a resistance equal to the output impedance of the lumped input with the noise source off would need to have to emit the same noise power into the network. Due to both the noise source and cryogenic switch impedances, this impedance is not perfectly matched to $50~\Omega$. As such, the extracted $T_\text{in}^\text{c}$ is higher than any physical temperature to account for reflections between the lumped input and DUT (or calibration attenuator).

For completeness, we also provide the mismatch-corrected equation used to extract $T_{50}$ for active HEMTs along with the correction coefficients in terms of measured S-parameters and reflection coefficients:

\vspace{-10pt}
\begin{align}
    \label{eq:T50Corr}
    T_{50} =&~ \frac{\mathbb{C}_1^hT_\text{in}^h-\mathbb{C}_1^cYT_\text{in}^c-\frac{T_{50}^\text{BE}(Y-1)}{|S_{21}^\text{DUT}|^2}}{\mathbb{C}_2^cY-\mathbb{C}_2^h}\\
    \mathbb{C}^\text{c(h)}_1 =&~ \frac{1-|\Gamma_{\text{in}}^\text{c(h)}|^2}{|1-S_{11}^\text{DUT}\Gamma_{\text{in}}^\text{c(h)}|^2|1-S_{11}^\text{BE}\Gamma_\text{out,DUT}^\text{c(h)}|^2}\\
    \mathbb{C}^\text{c(h)}_2 =&~ \frac{1-|\Gamma_\text{out,DUT}^{c(h)}|^2}{(1-|S_{22}^\text{DUT}|^2)|1-S_{11}^\text{BE}\Gamma_\text{out,DUT}^{c(h)}|^2}
\end{align}
 As is the usual case for the Y-factor technique without the use of an impedance tuner, this equation was derived by assuming that $T^\text{DUT}(\Gamma_\text{in}^\text{h})=T^\text{DUT}(\Gamma_\text{in}^\text{c})\equiv T_{50}$ and $T^\text{BE}(\Gamma_\text{out,DUT}^\text{h})=T^\text{BE}(\Gamma_\text{out,DUT}^\text{c})\equiv T_{50}^\text{BE}$. In other words, the noise power transfer between components is corrected for, but since the noise parameters of the DUT and backend are not known then their noise contributions are assumed to be unchanged between hot and cold impedances, which is not strictly true and introduces error.

Finally, we also rearrange \cref{eq:YFactAtten} to solve for the physical temperature of a passive component in thermal equilibrium:
\begin{equation}
    \label{eq:Tphys}
    T_\text{phys}=\frac{T_\text{in}^\text{h}G_\text{a}^\text{h}-\frac{Y}{\mathbb{C}}T_\text{in}^\text{c}G_\text{a}^\text{c}+T_\text{e}^\text{BE}(1-\frac{Y}{\mathbb{C}})}{\frac{Y}{\mathbb{C}}(1-G_\text{a}^\text{c})-(1-G_\text{a}^\text{h})}
\end{equation}

\begin{figure*}[t]
    {\phantomsubcaption\label{fig:PulsedNSdata}}
    {\phantomsubcaption\label{fig:Yfactordata}}
    \centering
    \includegraphics[width=0.90\textwidth]{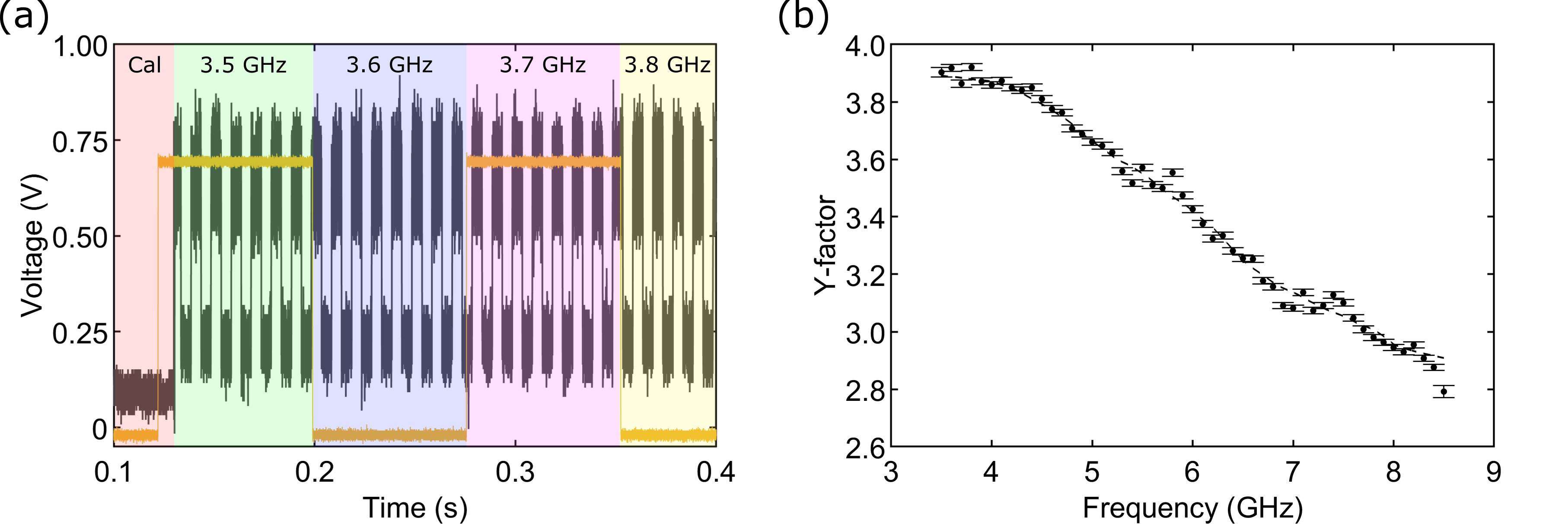}
    \caption{\textbf{(a) } Time series of the first 400~ms of a pulsed noise source, frequency swept Y-factor dataset for a calibration attenuator. Raw noise voltage data (black) shows noise source pulsing and trigger voltage (yellow) indicates a switch between swept parameter values, in this case frequency. Shown are the end of the calibration period for the first 0.12~ms, three frequency steps, and several hot/cold noise source pulses. Colored sections guide the eye between frequency steps. \textbf{(b)} Extracted Y-factor versus frequency. The dashed line is a guide for the eye.}
\end{figure*}

\section{Cryogenic probe station\label{sec:CPS} with pulse-biasing}
We now describe our cryogenic probe station (CPS) which we used to experimentally implement the calibration technique. Our CPS, which was originally reported in \cite{<russell_cryogenicprobe_2012>} and has been used extensively in a number of on-wafer transistor amplifier studies across the 1-115~GHz range \cite{varonen_75116-ghz_2013,cuadrado-calle_broadband_2017,kangaslahti_sub-20-k_2017,varonen_cryogenic_2020,choi_characterization_2021,gabritchidze_experimental_2022,gabritchidze_experimental_2024}, is shown schematically in \cref{fig:CPS}. For this study, it has received upgrades to perform pulsed cryogenic noise and S-parameter measurements, both on ungated (passive) and gated (active) HEMTs. A detailed schematic of the backend receiver is shown in \cref{fig:Backend}. A picture of the laboratory setup featuring key components is shown in \cref{fig:CPSpic}.

In order to switch in-vacuo between noise and S-parameter measurements, cryogenic SPDT latching switches (Radiall R570442000) were installed on the input and output paths inside the CPS. The switches were controlled by a relay board (SunFounder TS0012 8 Channel Relay) with a 12~V power supply and switched using a DAQ (NI USB-6259). Whereas previously two separate cooldowns were required to measure S-parameters and noise, these switches enable the full suite of DUT characterization to be performed without lifting the probe tips from the DUT pads. Since microwave measurements are sensitive to the probe-pad landing conditions \cite{mokhtari_new_2022}, this method enables more accurate and repeatable device characterization. A vector network analyzer (VNA; model Rhode\&Schwarz ZVA50) was used for S-parameter measurements, with a maximum calibrated sampling rate of $\sim70$~$\mu$s/sample. While an uncalibrated sampling rate down to $\sim3.5$~$\mu$s/sample was possible, and in fact pulsed S-parameter measurements are not strictly necessary in most cases since the DUT gain can be extracted from Y-factor measurements, the full S-parameter corrections are often needed for poorly matched devices whose impedance changes appreciably with bias, for example, in ungated HEMTs.

A high excess noise ratio (ENR) noise source (model MMW TECHNOLOGY INC 10~MHz-40~GHz NOISE SOURCE S/M12) was used to supply hot and cold noise for Y-factor measurements. It had an ENR of $\sim25$~dB at 5~GHz. We also built a fast noise source bias switch using a 28~V power supply and an IRFP140 power MOSFET to enable fast switching times up to $\sim5$~$\mu$s. Although we did not push this switching time to its limit set by the capacitance of the biasing circuitry, such fast switching times could, in principle, be used to take time domain noise measurements capable of resolving physical device heating dynamics on the order of $\sim$~$\mu$s.

The noise power was carried in K-connectorized (2.92~mm) coaxial cables. For HEMT measurements, the cold-attenuator Y-factor method was used\cite{gallego_accuracy_1992,fernandez_noise-temperature_1998} and a $\sim20$~dB cryogenic attenuator (Quantum Microwave QMC-CRYOATTK-20) was placed on the input line to achieve hot and cold noise powers on the order of tens to hundreds of kelvin to reach the DUT input. For ungated devices, the cold-attenuator was replaced with a cryogenic $0$~dB pad to thermalize the input line to the cold stage without attenuating the input noise. In this latter case, the hot and cold noise power reaching the DUT input were $>10,000$~K and $\approx300$~K, enabling noise temperatures in the thousands of Kelvin range to be resolved.***In this study, there are no ungated devices, so probably omit these last sentences.

We used $100$~$\mu$m pitch ground-signal-ground (GSG) DC--40~GHz probes (GGB Picoprobe 40A-GSG-100-DP) for the input (gate) and output (drain) DUT connections. S-parameter calibration was performed up to the probe tips using the SOLT method with GGB Picoprobe CS-5 calibration standards. On the output, another cryogenic switch was used to switch between the S-Parameter path and the noise path, which had a cryogenic LNA (Low Noise Factory LNF-LNC4\_8G) as the first stage amplifier followed by a room-temperature backend receiver shown in \cref{fig:Backend}. The numbered components are listed in \cref{tab:RTbackend}. The main features are: a mixer to downconvert the noise power, switches to convert between a narrow (4.4~MHz) and wide (197.5~MHz) signal path (to enable either accurate frequency dependent measurements or higher stability wideband measurements \cite{kooi_submillimeter_2017}), a 0-40~dB voltage-variable attenuator to enable measurement over a large range of input powers covering both the passive and active DUT regimes, and an ultrafast tunnel-diode detector with a Dicke-switched calibration load to enable zero-power voltage offset calibration of the detector before each measurement.

\begin{figure*}[t]
    {\phantomsubcaption\label{fig:PulsedBiasSpars}}
    {\phantomsubcaption\label{fig:PulsedBiasSparsZoom}}
    {\phantomsubcaption\label{fig:PulsedBiasHot}}
    {\phantomsubcaption\label{fig:PulsedBiasHotZoom}}
    \centering
    \includegraphics[width=0.90\textwidth]{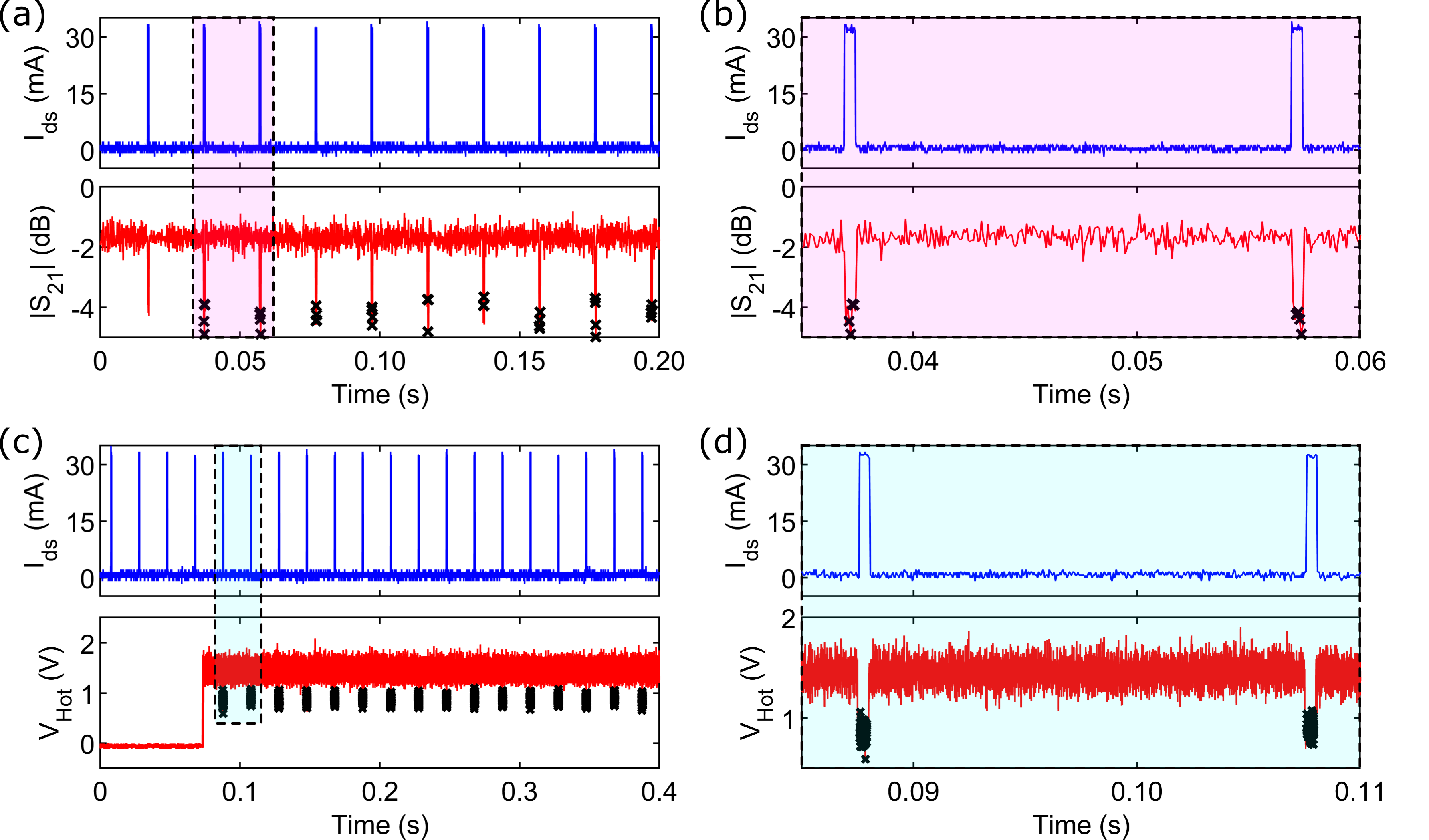}
    \caption{Representative $S_{21}$ and noise voltage time series measured on a gateless HEMT, illustrating the pulsed data extraction. \textbf{(a)} $S_{21}$ (red), corresponding current bias (blue), and extracted pulse-bias data points (black crosses). \textbf{(b)} Zoom on the dashed-line box enclosed region from \textbf{(a)}. \textbf{(c)} Hot noise voltage (red), corresponding current bias (blue), and extracted pulse-bias data points (black crosses). \textbf{(d)} Zoom on the dashed-line box enclosed region from \textbf{(c)}. Zoomed plots are shown to distinguish individual pulses.}
\end{figure*}

The detector diode output was sent through a preamplifier chain to a high-speed oscilloscope (Tektronix DPO72304DX 23 GHz Digital Phosphor Oscilloscope). Representative pulsed noise-source DC-bias data is shown in \cref{fig:PulsedNSdata}. The noise source was pulsed on and off at 100~Hz as the local oscillator frequency was swept from 3.5~GHz to 8.5~GHz in 0.1~GHz steps. Each voltage step shown in yellow represents the timing trigger supplied by a Keithley sourcemeter signaling the step between frequencies. All hot and cold voltages were averaged, respectively, and the ratio was taken to yield the Y-factor shown in \cref{fig:Yfactordata}. Error bars reflect the deviation in noise voltage averaged within each frequency step.

Pulse biasing was performed with an arbitrary waveform generator (AWG; model Agilent 33522A) to generate square pulses with fixed period and duty cycle. Current was measured using a high-speed current sense amplifier (Analog Devices AD8411AR, 2.5~MHz bandwidth), which amplifies the voltage dropped across a shunt resistor with a known, accurate resistance value, and is an inexpensive alternative to oscilloscope current probes with the additional ability to modify the current dynamic range by simply replacing the shunt resistor. A two-channel BNC-connectorized current-sense box was constructed to measure two different current ranges: 0-100~$\mu$A for measuring gate leakage current, and 0-200~mA for measuring drain current in both gated and ungated devices. The current-to-voltage transduction factor of each channel was calibrated by supplying a known current from a Keithley2400 sourcemeter through the shunt resistor in series with a fixed $50~\Omega$ load and measuring the final voltage on the oscilloscope.

\Cref{fig:PulsedBiasSpars} shows representative pulsed S-parameter and current data on an ungated HEMT. \Cref{fig:PulsedBiasHot} shows representative pulsed noise data with the noise source on. \Cref{fig:PulsedBiasSparsZoom,fig:PulsedBiasHotZoom} show zoomed plots to distinguish two individual pulses. Pulsing here was performed with a 20~ms period and 500~$\mu$s pulse width, for a duty cycle of 2.5~\%. The VNA sampling rate was set to its fastest calibrated rate of $\sim70$~$\mu$s/sample, and noise was sampled at 500~kHz. The digitization noise observed in the current data was caused by the current sense box outputting voltage from 0-5~V with $I=0$~A corresponding to $V=2.5$~V, and the oscilloscope used here was incapable of zeroing this voltage. The analog-to-digital converter in the scope limited the sensitivity to $\sim1$~mA. This is not an intrinsic limit of our measurement setup, and it can be remedied by modifying the the current sense box circuitry to output $V=0$~V at $I=0$~A. For the purposes of this paper, the current sensitivity of $\sim1$~mA is adequate.


\section{Results on InP HEMTs \label{sec:HEMTs}}

We now demonstrate our calibration technique by using it to measure the noise temperature ($T_{50}$) of a 4-finger 50~$\mu$m gate periphery InP HEMT (Diramics) which was previously measured in our CPS using a calibration technique where the noise source, coaxial lines, cold-attenuator, and backend receiver were independently characterized in separate cooldowns and their noise contributions cascaded. We compare the results of these measurements here. 

For the data set from our earlier measurements, the calibration was performed as follows. The S-Parameters and physical temperature of each component were measured independently and their noise contributions were cascaded assuming a linear temperature gradient along the cabling from room temperature to the cold-attenuator. The noise source was an Agilent N4001A 15~dB SNS whose ENR was determined using the factory-calibrated lookup table. The input line coaxial cabling loss was measured in a separate dewar as a function of stage temperature, to emulate the same thermal gradient as experienced in the CPS. The 20~dB cold-attenuator loss was also measured in a separate cooldown. The backend was calibrated using the Y-factor method landed on a thru, and using the previously calibrated the noise source and input lines. No corrections for component mismatch was used.

 \begin{figure*}[t]
    {\phantomsubcaption\label{fig:ColdAttenCalib}}
    {\phantomsubcaption\label{fig:DiramicsT50}}
    \centering
    \includegraphics[width=0.85\textwidth]{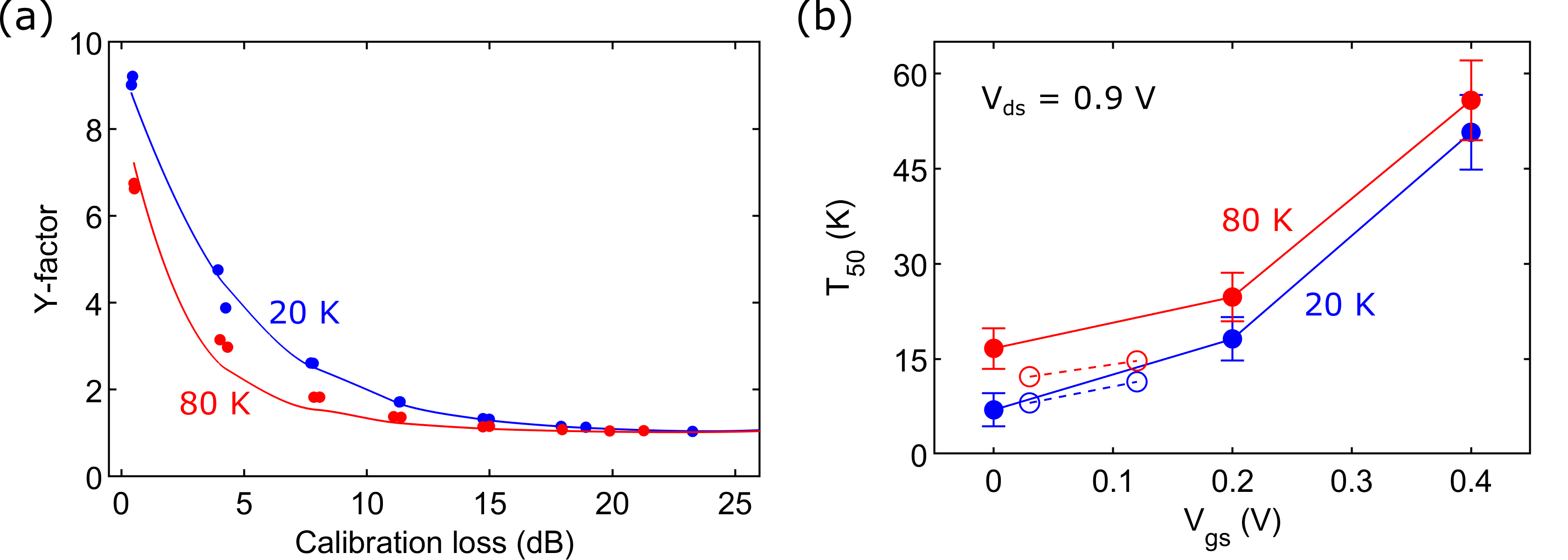}
    \caption{\textbf{(a)} Experimentally measured Y-factor vs attenuator loss with a 20~dB cold-attenuator on the input path. Plotting and extraction details were the same as in \cref{fig:CryoAttenExp}. \textbf{(b)} Noise temperature versus gate bias for a Diramics 4f50 discrete InP HEMT. Data was acquired at 20~K (solid blue circles) and 80~K (solid red circles). Previously acquired data which was calibrated using independently measured coaxial loss and estimated temperature gradients are also shown at 20~K (open blue squares) and 80~K (open red circles) for comparison.}
\end{figure*}

For the data set of our most recent measurements, a 20~dB cold-attenuator was inserted on the input line to reduce the magnitude of noise reaching the DUT to enable more precise noise temperature extraction. \Cref{fig:ColdAttenCalib} shows the Y-factor vs attenuator loss calibration data and lines of best fit. We extracted $T_\text{in}^\text{c}=23.1\pm1.4$~K, $T_\text{in}^\text{h}=272.7\pm19.0$~K, and $T_{50}^\text{BE}=24.2\pm4.3$~K, in reasonable agreement with estimates using the method described above. We first note that the fitting curves are smoother than in \cref{fig:CryoAttenExp} due to the improved impedance match on the input line when using the cold-attenuator. 

We also note that while the 20~K fitting curve appears to fit the data well, the 80~K fitting curve noticeable underestimates the data, a feature which was not observed in \cref{fig:CryoAttenExp}. We suspect this is primarily because the assumption that the input line physical temperature remains unchanged between the two physical stage temperatures is no longer an accurate approximation. From independent temperature measurements of the cold-attenuator using a DT-670 diode mounted to the cold-attenuator chassis, we measure a change in cold-attenuator temperature from 26~K to 35~K when changing stage temperature from 20~K to 80~K. While this temperature change was also observed in the 0~dB thermalization pad used in the ungated HEMT measurement setup, the loss was sufficiently low that this temperature change negligibly affected the noise temperature change between 20~K and 80~K, but temperature change of the lossy attenuator causes a more appreciable change in noise power.

This is clearly a limitation of this particular implementation of the calibration technique. The limitation could be overcome by employing a thermally isolated attenuator chip with on-chip heating and temperature sensing capabilities, as  mentioned previously.

\Cref{fig:DiramicsT50} shows the noise temperature of the Diramics device at a fixed drain-source bias $V_{ds}=0.9$~V at three gate-source biases. All data was taken at 5~GHz. Also shown are previously acquired noise temperature measurements of this device. While the agreement between separate measurements at 20~K is quantitatively good, our current calibration method overestimates the DUT noise relative to the previous measurements, although the disagreement is still within absolute uncertainty of each measurement. We expect that this is partially due to the change in cold-attenuator temperature mentioned above, and partially due to calibration uncertainties in our prior measurements which were not fully characterized.

\section{Conclusion} \label{sec:conclusion}

We have reported a cryogenic noise calibration technique that enables more accurate and repeatable noise temperature measurements of both passive and active on-wafer microwave semiconductor devices. This technique relies on the precise temperature and S-parameter measurements of a series of on-chip attenuators to yield the hot and cold power arriving at the input plane of the DUT and the backend noise temperature from the output probe tips. We validated our calibration technique both through comparison with simulations and with prior measurements on an InP HEMT. The method method can be further improved and extended to a wider range of physical temperatures by employing a module capable of switching between on-chip attenuators with adjustable temperature control.

\section{Acknowledgments}
The authors thank Bekari Gabritchidze, Jan Grahn, Pekka Kangaslahti, Jacob Kooi, and Junjie Li for useful discussions. We gratefully acknowledge the critical support and infrastructure provided for this work by The Kavli Nanoscience Institute at the California Institute of Technology. 

\bibliography{refs_AAZot_natbib}

\begin{IEEEbiographynophoto}{Anthony J. Ardizzi} received the HBSc degree in physics from the University of Toronto in 2015, and the M.S. and Ph.D. degree in applied physics from the California Institute of Technology in Pasadena, CA, in 2018 and 2022, respectively, specializing in precision microwave measurements and semiconductor noise characterization. He is currently a Research Engineer jointly with the Cahill Radio Astronomy Lab and the Minnich Lab, where he is actively researching both atomic-layer nanofabrication techniques applied to low-noise InP HEMTs and novel noise characterization techniques for microwave devices.
\end{IEEEbiographynophoto}

\begin{IEEEbiographynophoto}{Jiayin Zhang} received B.S. degree in mechanical engineering from the University of Rochester (UR), Rochester, NY, USA, in 2021. He is now pursuing Ph.D. degree in materials science at California Institute of Technology in Pasadena, CA, USA.
His research interests include high electron mobility transistor (HEMT) design and fabrication,  semiconductor noise characterization, and novel nanofabrication techniques.
\end{IEEEbiographynophoto}

\begin{IEEEbiographynophoto}{Akim A. Babenko} received the B.S. degree (Hons.) in electronics and nanoelectronics and the M.S. degree (Hons.) in radiophysics from Kuban State University, Krasnodar, Russia, in 2017 and 2019, respectively, with a focus on the vector network analysis of microwave mixers, and the Ph.D. degree in electrical and electronics engineering from Prof. Popovic’s Group, University of Colorado (CU) Boulder, Boulder, CO, USA, in 2022, for his work at the National Institute of Standards and Technology on the Quantum Voltage Project. He designed multidecade bandwidth passive and active MMICs in III–V and superconductor technologies and developed new measurement and signal-processing techniques for RF Josephson arbitrary waveform synthesizers. Since 2022, he has been with Jet Propulsion Laboratory, Pasadena, CA, USA, on low-noise and quantum-limited receivers for atmospheric remote sensing and radio astronomy.
\end{IEEEbiographynophoto}

\begin{IEEEbiographynophoto}{Kieran A. Cleary}received the M.Eng.Sc. degree in electronic engineering from The National University of Ireland, Dublin, Ireland, in 1994, and the Ph.D. degree in radio astronomy on cosmic microwave background observations using the Very Small Array from The University of Manchester, Manchester, U.K., in 2004. He is currently a Senior Scientist with the California Institute of Technology, Pasadena, CA, leading the CO Mapping Array Project (COMAP), a spectral line
intensity mapping experiment tracing the structure of the Universe in 3D. As lead of the Cahill Radio Astronomy Laboratory (CRAL), he works to advance low noise amplifiers and develop new instruments for radio astronomy. He is also an Associate Director of the Owens Valley Radio Observatory.
\end{IEEEbiographynophoto}

\begin{IEEEbiographynophoto}{Austin J. Minnich} is Deputy Division Chair of the Division of Engineering and Applied Science and a Professor in Engineering and Applied Science at the California Institute of Technology. He received his Bachelor’s degree from UC Berkeley in 2006 and his PhD from MIT in 2011, after which he started his position at Caltech. His research interests include low-noise microwave instrumentation, device physics, and atomic layer processing.

\end{IEEEbiographynophoto}

\end{document}